\documentclass{ws-p10x7-hepex}

\begin{document}

\title{ZZ Cross Section Measurements}

\author{Salvatore Mele}

\address{EP Division, CERN, CH-1211, Gen\`eve 23, Switzerland\footnote{}\\E-mail: Salvatore.Mele@cern.ch}

\twocolumn[\maketitle\abstract{
Results on the cross section measurement 
of Z boson pair--production at LEP\cite{ammucchiata,altri}  are presented. 
The more general case of neutral--current four--fermion 
production  and the particular case of ZZ events enriched in b quarks
are also discussed. All the results agree with the Standard Model
predictions.}] 

\section{Introduction}
\footnotetext{$^*$On leave of absence from INFN Sezione di Napoli, I-80126, Napoli, Italy.}

LEP was successfully operated in the years from 1997 through 2000 
at centre--of--mass ($\sqrt{s}$) energies from 183\,GeV up to
208\,GeV.
This allowed each of its four experiments to collect more than 540\,pb$^{-1}$
of data above the Z boson
pair--production threshold, as summarised in Table~\ref{tab:1}.

This process tests the Standard Model of electroweak
interactions 
in the neutral--current sector and is sensitive to New Physics scenarios
such as couplings between neutral gauge bosons\cite{matteuzzi}
or extra space dimensions\cite{gravitons}.

\begin{figure}
\epsfxsize200pt
\figurebox{120pt}{160pt}{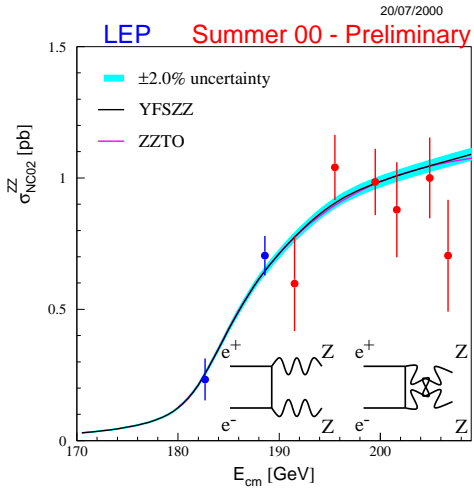}
\caption{Measured and predicted cross sections for the Z boson
  pair--production process via the two diagrams shown in the lower
  right corner.}
\label{fig:1}
\end{figure}

The results presented here refer to a particular subset of all the
possible diagrams for  neutral--current four--fermion production,
denoted as NC02 and  depicted in Figure~\ref{fig:1}. The Figure also shows
the ZZ production cross section as calculated with the YFSZZ and ZZTO
programs\cite{passarino}. These diagrams define the Z pair production signal,
also defined by some experiments with  a wider part of the full
four--fermion phase  space compatible with Z pair--production.

The related topics of general neutral--current four--fermion
production and b quark content in Z--pair events are also
investigated, and discussed in the following.

Detailed accounts of the data sets, 
analysis techniques and  results of each experiment can be found
elsewhere\cite{ammucchiata,altri}. All the results at
$\sqrt{s}=183$ and 189\,GeV are published
while the others are preliminary.

\section{Data analysis}

The four experiments devised analysis strategies that rely on
the identification of the signatures of the pair--production of two
particles with equal masses,  compatible with Z bosons.

Multivariate techniques are used for the largest
statistic fully hadronic final state (49\% of the Z pair decays), where
a large background is expected from the QCD and W pair--production
processes. 
Event shape variables are used to reject the
first, focusing  on the signal four--jet topology,
also common to hadronic decays of W pairs. This background can be
discriminated thanks to  the 
different boson mass that reflects into 
the topology  of the decay products. Moreover,
W decays lack b quarks, present
in 39\% of the fully hadronic decays 
of the Z pairs. The use of the b tag techniques developed 
for the Higgs search hence increases the signal purity. 

The final state with a Z decaying into 
hadrons and the other into neutrinos is the second most populated
(28\%) and is investigated 
with hadronic events with large missing energy. The 
hermeticity of the detectors allows to reconstruct the four--momentum
of the Z decaying into neutrinos. The main backgrounds are the
production of a Z  in association with an undetected high energy and
low polar angle initial state radiation photon, semileptonic decay of
W pairs with the charged lepton escaping detection or the more general
case of a W decaying into hadrons produced together with  a non
resonant system of a low polar angle electron and a
neutrino. Sequential cuts or multivariate 
techniques that enforce the signal topology of an undetected Z suppress
these backgrounds.

High signal purity is achieved in the lower statistic (14\%)
final state with hadrons and charged  leptons, which benefits of 
the high resolution measurements of the lepton momenta.
Kinematic fits in the hypothesis of an equal mass of the lepton--lepton
and hadron systems are performed, requiring then this mass to be
compatible with the Z
mass. Topological variables such as the angles 
of the leptons and the hadronic jets are also used.

The final states with two charged leptons and two neutrinos (4\%) and
four charged leptons (1\%) are penalised by low statistics, even though
identified with good purity thanks to the high lepton 
resolutions. The background from lepton pair--production and
four--fermion processes is mainly rejected by requiring the lepton
invariant and recoil masses  to
be compatible with the Z mass.

The undetectable final state with four neutrinos accounts only for 4\%
of the Z--pair decays. 

\section{Results}

The four experiments measured the
Z pair-production cross section at all the
$\sqrt{s}$ above threshold\cite{ammucchiata,altri}.
Combined results  are presented in
Table~\ref{tab:1} and Figure~\ref{fig:1}. This average takes into
account sources of common and correlated
systematic uncertainties, mainly due to uncertainties on the
background cross sections and modelling, as well as uncorrelated ones,
dominated by Monte Carlo statistics and 
detector related effects, in particular for the b tag
procedure. 

\begin{table}[t]
\caption{LEP $\sqrt{s}$ and sum of the integrated luminosities (${\cal{L}}$)
collected by the four experiments together with the
measured ($\sigma^{\rm ZZ\,\,Exp}_{\rm NC02}$) and expected
($\sigma^{\rm ZZ\,\,Th}_{\rm NC02}$) ZZ cross sections. 
Data collected in the year 2000 are grouped into the two last
energy bins. The first uncertainty is statistical, the second
  systematic.} 
\label{tab:1}
\begin{tabular}{|c|c|c|c|c|}  
\hline
$\sqrt{s}$ & ${\cal{L}}$&$\sigma^{\rm ZZ\,\,Exp}_{\rm NC02}$ & 
$\sigma^{\rm ZZ\,\,Th}_{\rm NC02}$ \\
GeV & pb$^{-1}$ &pb & pb \\
\hline
182.7 & 221 &$0.23 \pm 0.08 \pm 0.02$ & 0.25 \\
188.7 & 686 &$0.70 \pm 0.07 \pm 0.03$ & 0.65 \\
191.5 & 114 &$0.60 \pm 0.18 \pm 0.04$ & 0.77 \\
195.6 & 310 &$1.04 \pm 0.12 \pm 0.04$ & 0.90 \\
199.6 & 326 &$0.98 \pm 0.12 \pm 0.04$ & 0.98 \\
201.7 & 152 &$0.88 \pm 0.18 \pm 0.04$ & 1.01 \\
205.0 & 241 &$1.00 \pm 0.15 \pm 0.05$ & 1.04 \\
206.8 & 123 &$0.70 \pm 0.21 \pm 0.05$ & 1.06 \\
\hline
\end{tabular}
\end{table}

All the measurements agree with the Standard Model predictions.
The ratio between the measured and
predicted cross sections is formed at each $\sqrt{s}$, and its average
over all the $\sqrt{s}$  
yields $0.99\pm0.06$, what reveals an overall  agreement within the
combined accuracy of 6\%, mainly statistical. 

\section{Four--fermion production}

The DELPHI Collaboration extends its Z pair--production analysis to the case
of a
Z boson produced in association with a pair of fermions from a
virtual photon. Final states with two quarks and
either a muon-- or a 
neutrino--pair are analysed. In the latter case
the mass of the hadronic system must be below 60\,GeV. The
cross sections over the full data sample are respectively expected to be 
$0.19-0.25$\,pb
and $0.13-0.16$\,pb, decreasing with $\sqrt{s}$. The measured cross
sections read:
\begin{displaymath}
\sigma^{183\,{\rm GeV}-208\,{\rm GeV}}_{\rm e^+e^-\rightarrow Z
  \gamma^\star \rightarrow \mu^+\mu^-q\bar{q}}  
= 0.22 \pm 0.05 \pm 0.02\rm\,pb 
\end{displaymath}
\begin{displaymath}
\sigma^{183\,{\rm GeV}-208\,{\rm GeV}}_{\rm e^+e^-\rightarrow Z
  \gamma^\star \rightarrow  \nu\bar{\nu}q\bar{q}} 
= 0.19 \pm 0.06 \pm 0.02\rm\,pb\,,
\end{displaymath}
where the first uncertainty is statistical and the second systematic.

\section{B quark content}

The experimental investigation of Z--pair final states containing b
quarks validates the capability of the LEP 
experiments to detect the production and decay of the Higgs boson.

\begin{figure}
\epsfxsize200pt
\figurebox{120pt}{160pt}{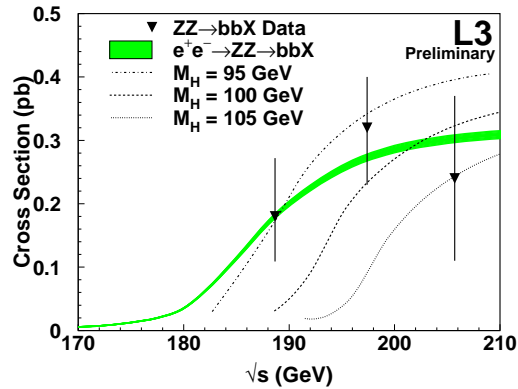}
\caption{Measured and predicted cross sections for neutral--current
  four--fermion events compatible with Z pairs with at
  least a b quark pair. Data are presented for the average
  $\sqrt{s}$ of the years 1998, 1999 and 2000. The
  computed cross section for the production of 
  the Standard Model Higgs boson is also presented for different mass
  hypotheses.}
\label{fig:2}
\end{figure}

These two process have a similar topology, as the Higgs boson
production would
preferentially manifests as a pair of b quarks from the
decay of a heavy object that recoils against a Z. The expected
cross sections are also similar, as illustrated in
Figure~\ref{fig:2}. The Figure also shows the results of the
measurement by the L3 Collaboration, in agreement with the
Standard Model predictions. In this measurement Z decays into hadrons,
neutrinos and charged leptons are considered in association with the b
quark pair.

The OPAL Collaboration measures the branching
ratio of the Z into b quarks in the selected Z--pair events, als
finding agreement with the 
value measured at LEP at the Z resonance. 

In conclusion, if such rare processes with these topologies can be
observed, a Higgs boson light enough to be produced at LEP will not escape
detection. 

\section*{Acknowledgements}

I wish to thank my L3 and LEP colleagues working on these subjects 
for all the constructive discussions we had throughout the last years
and for sharing with me their preliminary results.

\end{document}